# All-angle Negative Reflection with An Ultrathin Acoustic Gradient Metasurface: Floquet-Bloch Modes Perspective and Experimental Verification

**Bingyi Liu[1], Jiajun Zhao[2], Xiaodong Xu[1], Wenyu Zhao[1,3] and Yongyuan Jiang[1,4,5,6,*]**

*1 Institute of Modern Optics, Department of Physics, Harbin Institute of Technology, Harbin 150001, China*

*2 GOWell International LLC, Houston, Texas 77041, United States*

*3 Department of Physics, University of California at Berkeley, Berkeley, California 94720, United States*

*4 Key Lab of Micro-Optics and Photonic Technology of Heilongjiang Province, Harbin 150001, China*

*5 Key Laboratory of Micro-Nano Optoelectronic Information System of Ministry of Industry and Information Technology, Harbin 150001, China*

*6 Collaborative Innovation Center of Extreme Optics, Taiyuan 030006, Shanxi, People's Republic of China*

*Corresponding author: jiangyy@hit.edu.cn

## Abstract

Metasurface with gradient phase response offers new alternative for steering the propagation of waves. Conventional Snell's law has been revised by taking the contribution of local phase gradient into account. However, the requirement of momentum matching along the metasurface sets its nontrivial beam manipulation functionality within a limited-angle incidence. In this work, we theoretically and experimentally demonstrate that the acoustic gradient metasurface supports the negative reflection for all-angle incidence. The mode expansion theory is developed to help understand how the gradient metasurface tailors the incident beams, and the all-angle negative reflection occurs when the first negative order Floquet-Bloch mode dominates inside the metasurface slab. The coiling-up space structures are utilized to build desired acoustic gradient metasurface, and the all-angle negative reflections have been perfectly verified by experimental measurements. Our work offers the Floquet-Bloch modes perspective for qualitatively understanding the reflection behaviors of the acoustic gradient metasurface, and the all-angle negative reflection characteristic possessed by acoustic gradient metasurface could enable a new degree of the acoustic wave manipulating and be applied in the functional diffractive acoustic elements, such as the all-angle acoustic back reflector.

Metasurfaces, the quasi 2D metamaterials composed by elaborately arranged artificial scatters of subwavelength geometrical size, have shown powerful wavefront manipulation capabilities over the past few years. Based on the idea of sampling the desired wavefront and replacing the pixels with appropriate nanostructures[1,2], which exactly behave like the secondary sources proposed in Huygens' principle, optical metasurfaces constructed with plasmonic nano-antenna of various geometrical shape or low loss dielectric nano-post have been experimentally demonstrated to function as the ultrathin planar lenses[3,4], holograms[5,6] and low profile conformal optical devices[7-9]. As another important form of classical wave, acoustic wave can also be flexibly tailored by deep subwavelength diffraction inclusions, which is known as the acoustic metasurfaces[10-12]. Following the strategy that tunes the effective refractive index by coiling–up spaces[13-17], patterning the surface impedance profiles of metasurfaces[18,19] or taking advantage of low/high quality factor resonators[20-22], numerous types of acoustic metasurface structures have been utilized to build up functional acoustic lens[10,16,23,24], acoustic vortices generator[25,26], acoustic wave absorber[27,28] and ultrathin acoustic ground cloak[29,30].

Gradient metasurface composed by periodic supercells has been well studied for it steers the incident waves in the anomalous way governed by the generalized Snell's law[1,31-33]. The nature of momentum matching dictates that the contribution of local phase gradient cannot be ignored when the incident beam encounters the gradient metasurface. Therefore, the incident beam would be deflected asymmetrically under the all-angle illumination. Furthermore, when the beam is incident beyond the critical angle, no free space scattered fields exist, and only the non-propagating evanescent field, which is termed as the surface bounded waves[12,31], can be obtained. Notwithstanding that, recently, several research articles have reported that the apparent free-space propagating scattered field, which is featured as negative reflection[34,35] and negative refraction[12,36,37], can be observed even when the beam illuminates beyond the critical angle. These intriguing phenomena indicate that the underlying physical mechanism of the beam manipulation functionality of the acoustic gradient metasurface is still needed to be studied.

Similar to the electromagnetic gradient metasurface[31,38], in this work, we develop the mode expansion theory to help study and understand the mechanism of the extraordinary negative reflection behaviors of the acoustic gradient metasurface when it is illuminated under all-angle incidence. The acoustic gradient metasurface studied in this work can be treated as the waveguide with periodical parameter modulation. According to the Floquet theorem[39-41], such gradient metasurface could support a series of Floquet-Bloch modes with different propagation wave vectors. These Floquet-Bloch modes collectively contribute to the free space propagating field via energy transfer with different diffraction orders. Similar to the wide-angle negative reflection taking place at the free space-phononic crystal interface[42], when the first negative order Floquet-Bloch mode dominates, the negative reflection can be obtained. In order to verify the all-angle negative reflection, we use the coiling-up space structures to construct the desired reflected acoustic metasurface and experimentally study its scattered characteristics. The perspective of Floquet-Bloch mode for all-angle negative reflection has been well validated by both the theoretical calculations and experimental measurement results, and the apparent negative reflection indeed exists beyond the critical angle incidence.

## Results

**Floquet-Bloch modes excited inside the acoustic gradient metasurface.** The unique characteristic of all-angle negative reflection is that both the incident wave and reflected wave appear at the same side of the normal line for all-angle incidence[34], see Figure 1(b). Such nontrivial phenomenon can be regarded as the counterpart of the negative refraction occurring at the interface of two dissimilar medium with opposite refractive index values, which is a typical application of negative refractive index metamaterials, see Figure 1(a). In this work, we will introduce the perspective of Floquet-Bloch mode to help understand how the gradient metasurface interacts with the incident field, and how the acoustic energy trapped by the metasurface slab is finally tunneled into the $-1$ order diffraction for all-angle incidence.

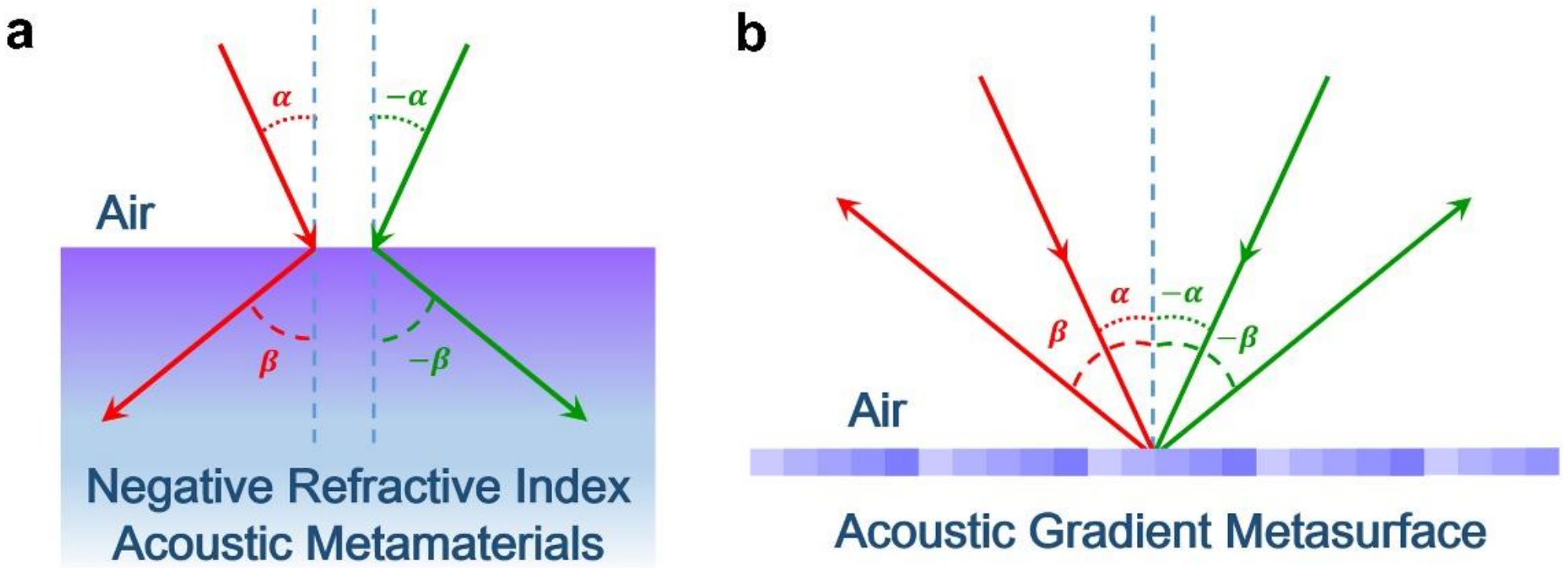

**Figure 1 All-angle negative reflection with an ultrathin acoustic gradient metasurface.** (a) Negative refraction occurring at the interface between air and negative refractive index metamaterials. (b) All-angle negative reflection realized by an ultrathin acoustic gradient surface.

The acoustic gradient metasurface(AGM) we study here is modeled as an anisotropic waveguide sandwiched between the free space(air) and the rigid wall(sound hard boundary). Figure 2(a) schematically illustrates the model system with continuous parameter distributions, the studying area is divided into three regions, I, II and III, which correspond to the free space, metasurface slab and rigid wall respectively. The density and bulk modulus of metasurface are generally given as

$$\vec{\rho}(x)=\rho_0\begin{pmatrix}\rho_x(x) & 0 & 0\\ 0 & \rho_y(x) & 0\\ 0 & 0 & \rho_z(x)\end{pmatrix},\quad \vec{B}(x)=B_0\begin{pmatrix}B_x(x) & 0 & 0\\ 0 & B_y(x) & 0\\ 0 & 0 & B_z(x)\end{pmatrix},\quad (1)$$

where $\rho_0$ and $B_0$ refer to the fluid density and fluid bulk modulus of air. Since the thickness of metasurface is of deep subwavelength scale, the parameter variation along the $z$ direction can be neglected. To simplify the study, we further assume that the metasurface slab is only inhomogeneous along $x$ direction while $y$ and $z$ direction are invariant, i.e., $\rho_y(x)=\rho_z(x)=\rho_x(x)$ and $B_y(x)=B_z(x)=B_x(x)$. Therefore, the underlying mechanism of all-angle negative reflection phenomenon can be well elucidated by comprehensively studying the scattering properties of this inhomogeneous system.

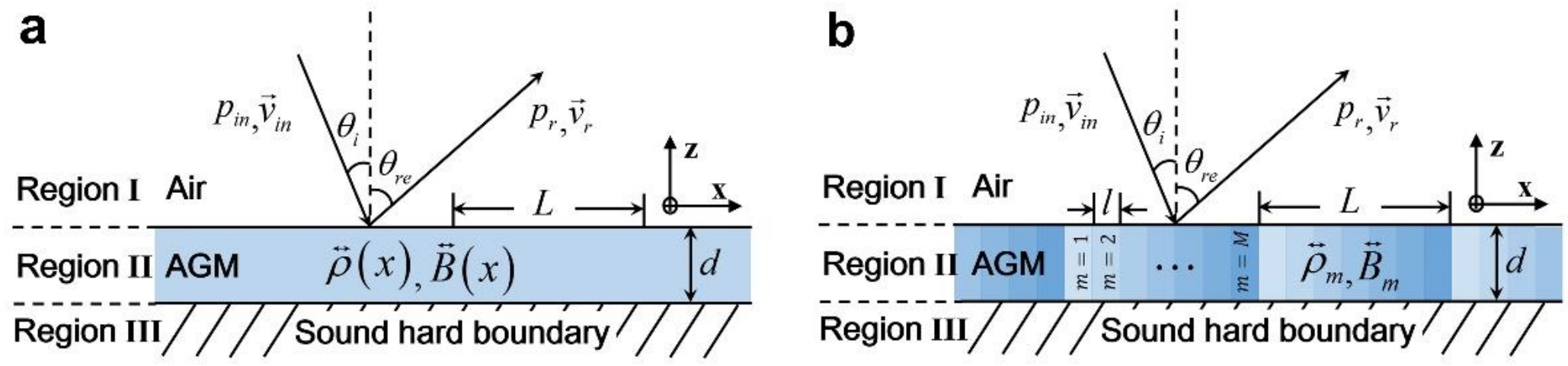

**Figure 2 Continuous and discretized model for analytically studying the acoustic gradient metasurface.** (a) Geometry of the acoustic gradient metasurface(AGM) slab with continuous parameter distribution, here $d$ is the thickness of the metasurface slab and $L$ is the length of the modulation period. (b) The discretized model of the acoustic gradient metasurface utilized for real practice.

When acoustic gradient metasurface is illuminated by the free space propagating plane waves, the metasurface subunits would strongly interact with the incident acoustic waves and reradiate part of the acoustic energy captured by the metasurface. As a waveguide, the intrinsic waveguide modes would be excited inside the metasurface slab, and these excited waveguide modes then couple the energy into different diffraction channels at the air-metasurface interface which forms the free space scattered fields.

In region I, the acoustic pressure field $p^I$ can be expanded as a series of plane waves (Rayleigh expansion) of the form

$$p^I = p^{in} + \sum_n R_n p_n^r = p^{in} + \sum_n R_n \exp\left(ik_x^{r,n} x\right) \exp\left[i\left(k_z^{r,n} z - \omega t\right)\right], \tag{2}$$

here $p^{in}$ is the acoustic plane wave incident at $\theta_i$, $p_n^r$ is $n$-th order reflected wave and $R_n$ is the corresponding reflection coefficient. Moreover, the parallel wave vector components of incident and $n$-th order reflected wave are defined as $k_{in} = k_0 \sin\theta_i$ and $k_x^{r,n} = k_0 \sin\theta_i + nG\left(n = 0, \pm 1, \cdots\right)$, where $k_0 = \omega / c_0$ is the amplitude of wave vector in air, $\omega$ is the angular frequency, $c_0$ is the sound speed of air, $G = 2\pi / L$ is the amplitude of reciprocal lattice vector.

Inside the region II, the density and bulk modulus of metasurface slab are functions of $x$ positions, the corresponding pressure and velocity field can be solved by relation $\vec{\rho}\frac{\partial \vec{v}}{\partial t} = -\nabla p$ and $\frac{\partial p}{\partial t} = -\nabla \cdot \vec{B}\vec{v}$, which deduce

$$\frac{B_z}{\rho_z}\frac{\partial^2 p^{II}}{\partial z^2}+\frac{\partial}{\partial x}\left(\frac{B_x}{\rho_x}\frac{\partial p^{II}}{\partial x}\right)+\frac{\rho_0}{B_0}\omega^2 p^{II}=0, \tag{3}$$

here $p^{II}$ refers to the acoustic field in the metasurface slab. For a periodic system, the Floquet theorem dictates that the steady propagation modes would be modulated by the supercell periodicity, which are in the form of Floquet-Bloch waves. Therefore, omitting the time dependent term $e^{-i\omega t}$, we can rewrite the acoustic pressure field inside the metasurface slab( region II) as

$$p_{\pm}^{II}(q_z,x,z)=\sum_n G(q_{z,n},x)\mathrm{e}^{\mp iq_{z,n}z}=\sum_n\left[h_{1,n}g^{+}(q_{z,n},x)+h_{2,n}g^{-}(q_{z,n},x)\right]\mathrm{e}^{\mp iq_{z,n}z}, \tag{4}$$

here $g^{+}(q_{z,n},x)$ and $g^{-}(q_{z,n},x)$ are two pseudo periodic functions satisfying $g^{\pm}(q_{z,n},x+L)=g^{\pm}(q_{z,n},x)\exp(-ik_0\sin\theta_i L)$, which represent the $n$-th order right-going (along the positive direction of $x$ axes) and left-going (along the negative direction of $x$ axes) Floquet-Bloch modes respectively, $h_{1,n}$ and $h_{2,n}$ are corresponding complex amplitude. The superscript sign $\mp$ of term $\mathrm{e}^{\mp iq_{z,n}z}$ stands for the forward($-$) and backward($+$) components of the field, which correspond to the acoustic energy coupling into and out of the metasurface slab respectively. Once the parameter distribution of the metasurface slab is known, the field inside the metasurface can be solved based on equation (3) and the Floquet-Bloch boundary condition at the two sides of one supercell[39-41].

Here we take the acoustic gradient metasurface with $k_0$ surface phase gradient as an example. When plane acoustic wave illuminates the acoustic gradient metasurface, several Floquet-Bloch modes can be excited inside the metasurface slab. Since the high order Floquet-Bloch modes possess greater average propagation constants, these modes are all evanescent and do not contribute to the free space scattered fields, therefore, only the first Floquet-Bloch modes should be taken into account. Owing to the local phase gradient, the excitation of the first Floquet-Bloch is propagating direction dependent, which means the Floquet-Bloch mode propagating along the direction of surface phase gradient is always the priority. Here we define the negative orders as the Floquet-Bloch modes propagate against the direction of the parallel wave vector component of the incident wave. When $\theta_i<0$, the $-1$ order Floquet-Bloch mode $g^{+}(q_{z,1},x)$ is excited and forms the negative reflections. When $\theta_i>0$, the former Floquet-Bloch mode $g^{+}\left(q_{z,1},x\right)$ would turn to be evanescent and forms the surface guided mode, then the acoustic energy of surface guided mode would be tunneled into the $-1$ order Floquet-Bloch mode $g^{-}\left(q_{z,1},x\right)$ via Bragg scattering of the supercells, which finally contributes to negative reflections.

Considering the realization of acoustic gradient metasurface for real practice, the discretized model depicted in Figure 2(b) is utilized for further analysis. In this case, one supercell is composed by *M* subunits and the width of each subunit is $l = L/M$. The fluid density and bulk modulus of the *m*-th subunit are denoted as $\vec{\rho}_m$ and $\vec{B}_m$, which can be regarded as two constants within the interval $\left(m\left(l-1\right),ml\right)$. Therefore, for the *m*-th subunit, equation (3) would be reduced into a typical wave equation, because the second term of equation (3) becomes zero. The general solution of such wave equation is a superposition of two plane waves with different amplitudes propagating in the opposite directions. Moreover, we need to determine the equivalent boundary condition connecting the adjacent subunits for further rigorous analytical calculations. In principle, matching the acoustic pressure field and z-component of the velocity at the boundaries of $z=0$ and $z=-d$, we are able to completely solve the reflection coefficient, and the scattering properties of the acoustic gradient metasurface can be well understood. More detailed information about the mode expansion theory for acoustic gradient metasurface can be found in the supplementary materials.

**Unit design, sample fabrication and experiment setup.** In this work, we utilize the coiling-up space structures proposed by *Li* et al[24] as the building blocks for acoustic gradient metasurface. Such structure can be regarded as a coiling channel with one end sealed by hard wall, and the acoustic wave propagates inside such deep-subwavelength channel would follow a zigzag footpath[16]. The metasurface is designed to operate at 2500 Hz and Figure 3(a) is a schematic illustration of one sample subunit, here the width *p* is 27 mm, the height *d* is 27 mm, the thickness of the horizonal bar *w* is 1.5 mm. Figure 3(c) shows the simulation result of the scattered field of five subunits with gradient phase shift ranging from 0 to 2π. The metasurface sample is fabricated with photosensitive resin via stereolithography (SLA, 0.08 mm in precision) and Figure 3(b) shows a fabricated supercell consists of 5 subunits presented in Figure 3(c), which is designed to realize the $-k_0$ surface phase gradient.

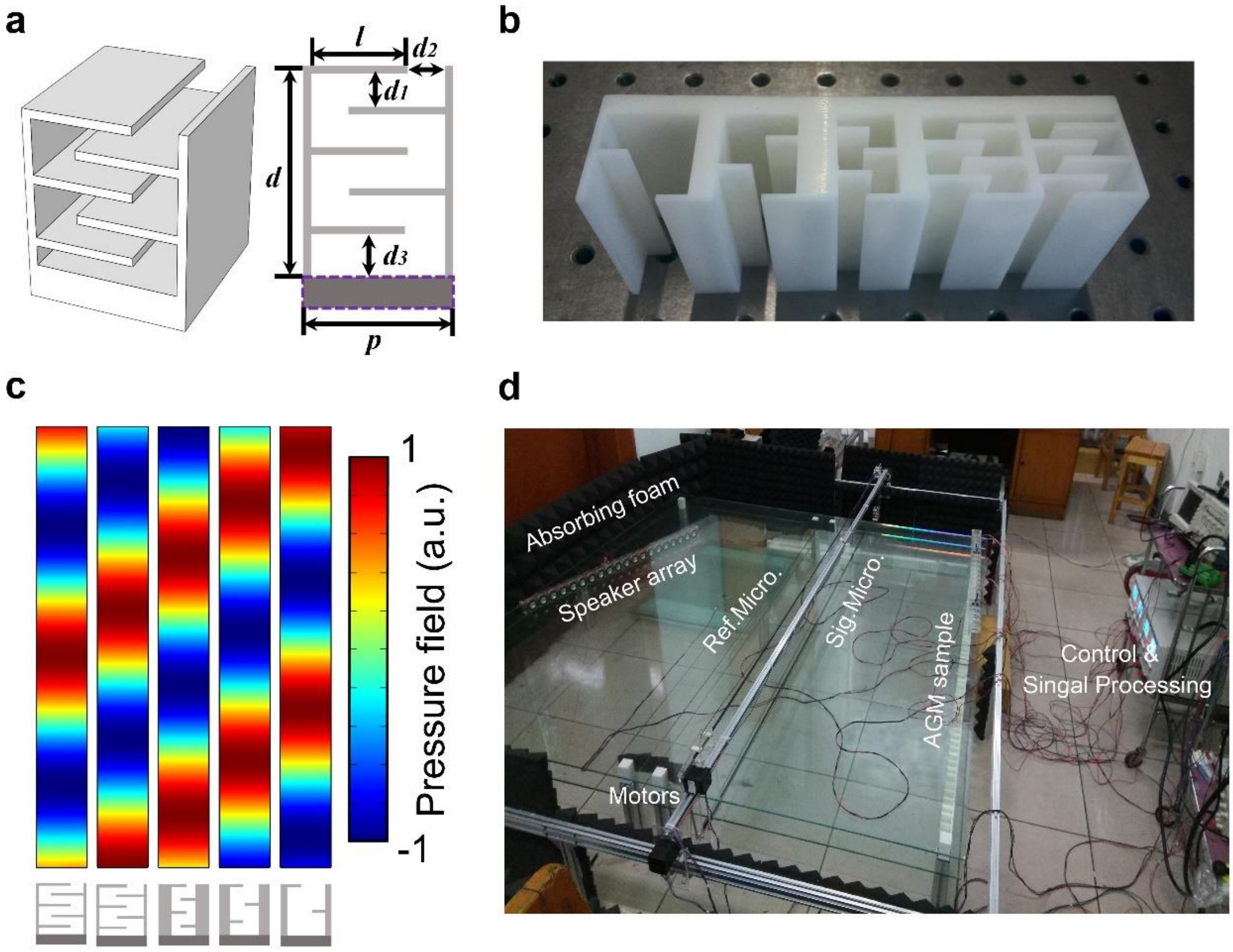

**Figure 3 Designed coiling-up space units and experiment setup.** (a) Schematic illustration of the coiling-up space structure and some parameters we utilized for the subunit design. The dark color region enclosed by purple dash line is the thick bottom(6 mm for real application) to mimic the sound hard boundary. (b) The fabricated supercell sample. (c) Scattered acoustic field of the coiling-up space structures utilized for sample fabrication possess the phase response covering $2\pi$ range. (d) Experimental 2D scanning system for full-field measurements.

In the experiment, the metasurface is composed by 9 supercells to mimic the infinite period system utilized in theoretical study. Then the beam reflection properties of this gradient metasurface via the home-built 2D acoustic waveguide is studied and the data are collected through a matched 2D-scanning measurement system. The metasurface sample is sandwiched between two pieces of tempered glass(1.9 m length, 1.8 m width, 7 mm thick) and the gap is chosen as 5 cm, of which the cutoff frequency is higher than the operating frequency. Here, the incident acoustic wave is generated by 20 tightly arrayed loudspeakers, and the acoustic plane wave can be obtained inside the waveguide. Two microphones are utilized to measure the acoustic pressure field, one microphone is fixed as the reference channel and another one working as the signal channel is attached to the synchronous belt driven by a stepper motor. Figure 3(d) depicts the layout of the experimental system.

**Experimental validation of all-angle negative reflection.** It is well known that the gradient metasurface deflects the incident plane waves in the way that governed by the generalized Snell's law. Furthermore, the generalized Snell's law only permits the negative reflection within a limited incident angle illumination, which is different from the all-angle negative reflection illustrated in Figure 1(b). However, the negative reflection occurring beyond the critical angle indeed exists when the first negative Floquet-Bloch mode is efficiently excited within the metasurface. Therefore, the gradient metasurface slabs composed by different artificial inclusions may function great differently for their distinct mode response properties, which should account for the high efficiency propagating wave-surface wave conversion operating in microwave regime[31] while the apparent negative reflection can be realized with the acoustic gradient metasurface. For the acoustic gradient metasurface demonstrated in Figure 2(a) and (b), the typical reflected phase shift is governed by $\varphi(x)=\varphi_0+\xi x$, here, $\varphi_0$ is determined by the average impedance value of metasurface and $\xi$ refers to the surface phase gradient. Based on the subunits presented in Figure 3(b), we theoretically study the scattering properties of the acoustic gradient metasurface with surface phase gradient $-k_0$. Figure 4(c) and (d) demonstrate the calculated scattered field when the gradient metasurface is illuminated under $\pm 30^\circ$ incidence. It is obvious that the typical apparent negative reflection occurred beyond the critical angle can be observed. Figure 4(e) illustrates the relation between the incident angle $\theta_i$ and reflected angle $\theta_{re}$ which follows the trajectory predicted by

$$\left(\sin\theta_{re}-\sin\theta_i\right)k_0=-\mathrm{sgn}\left(\theta_i\right)\xi\,, \tag{5}$$

here $\mathrm{sgn}\left(\theta_i\right)$ denotes the sign of incident angle and $\xi=-k_0$ . For gradient metasurface, the amplitude of surface phase gradient exactly equals to the reciprocal lattice vector. Therefore, the abnormal beam deflections governed by the generalized Snell's law can be understood as the Floquet-Bloch mode possessing the propagation constant $\vec{k}_{in}+\vec{\xi}$ couples the energy into the free space diffraction order.

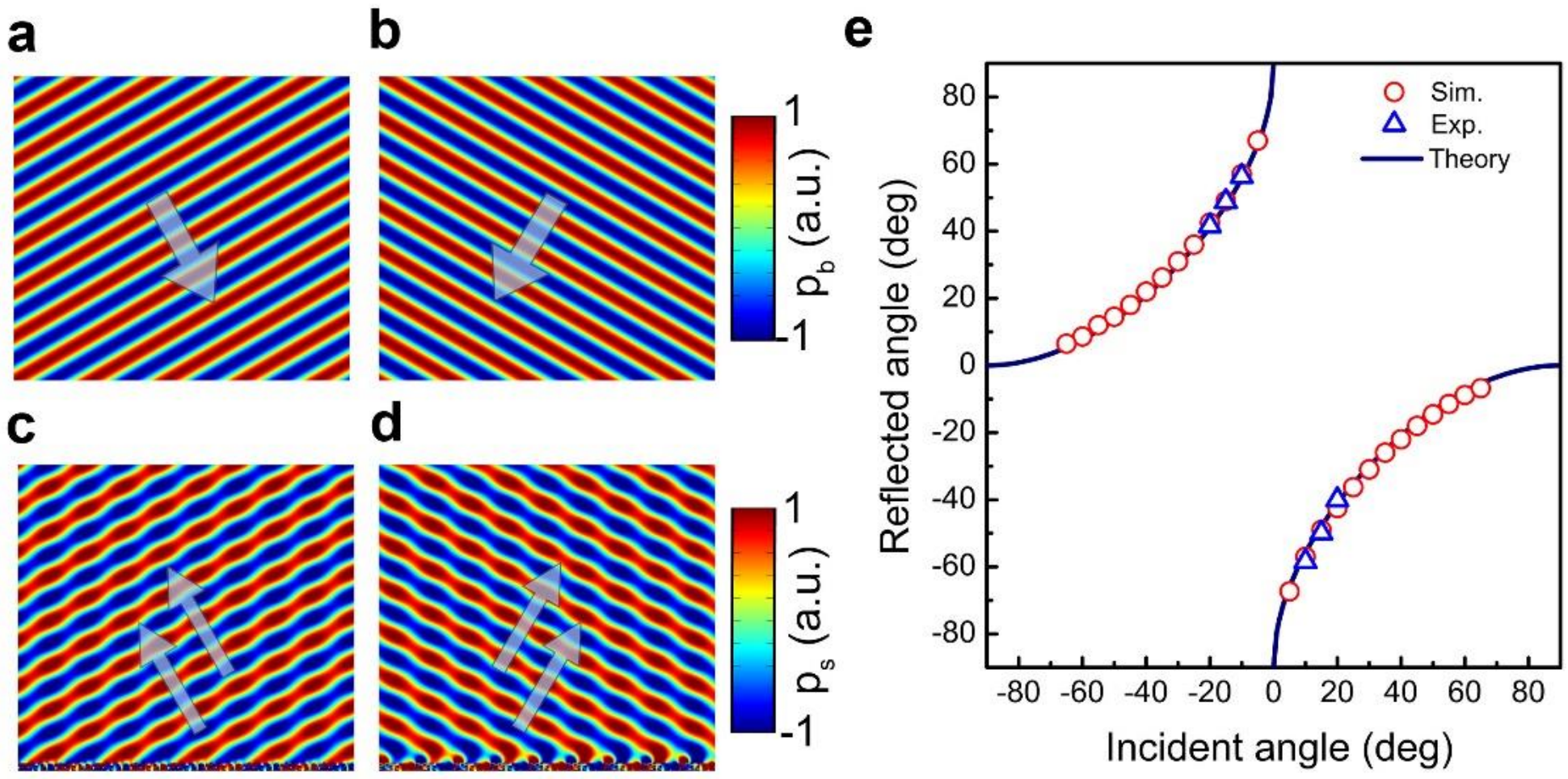

**Figure 4 Scattering properties of the acoustic gradient metasurface with $-k_0$ surface phase gradient.** The plane acoustic wave incident at (a) $30^\circ$ and (b) $-30^\circ$. The calculated scattered field of the AGM reflects at (c) $-31^\circ$ ($30^\circ$ incidence) and (d) reflects at $31^\circ$ ($-30^\circ$ incidence). (e) The relation between incident angle $\theta_i$ and reflected angle $\theta_{re}$. The solid lines represent the theoretical values, the red hollow circles stand for the simulated results and the blue hollow triangles are experimental data.

To verify the negative reflection occurred beyond the critical angle incidence, we construct a home-built 2D acoustic pressure field scanning measurement system, as shown in Figure 3(d). In the experiment, we simplify the measuring process by inversing the direction of the gradient metasurface to obtain an opposite surface phase gradient value instead of inversing the value of incident angle, therefore, we only need to measure the background field once to obtain two sets of scattered fields. Then we measure the scattered field of the gradient metasurface with $k_0$ or $-k_0$ surface phase gradients under $-10^\circ$, $-15^\circ$ and $-20^\circ$ incident angle illuminations respectively, which is exactly equivalent to the situation that the acoustic gradient metasurface having $k_0$ surface phase gradient being illuminated under $\pm 10^\circ$, $\pm 15^\circ$ and $\pm 20^\circ$ incidence correspondingly.

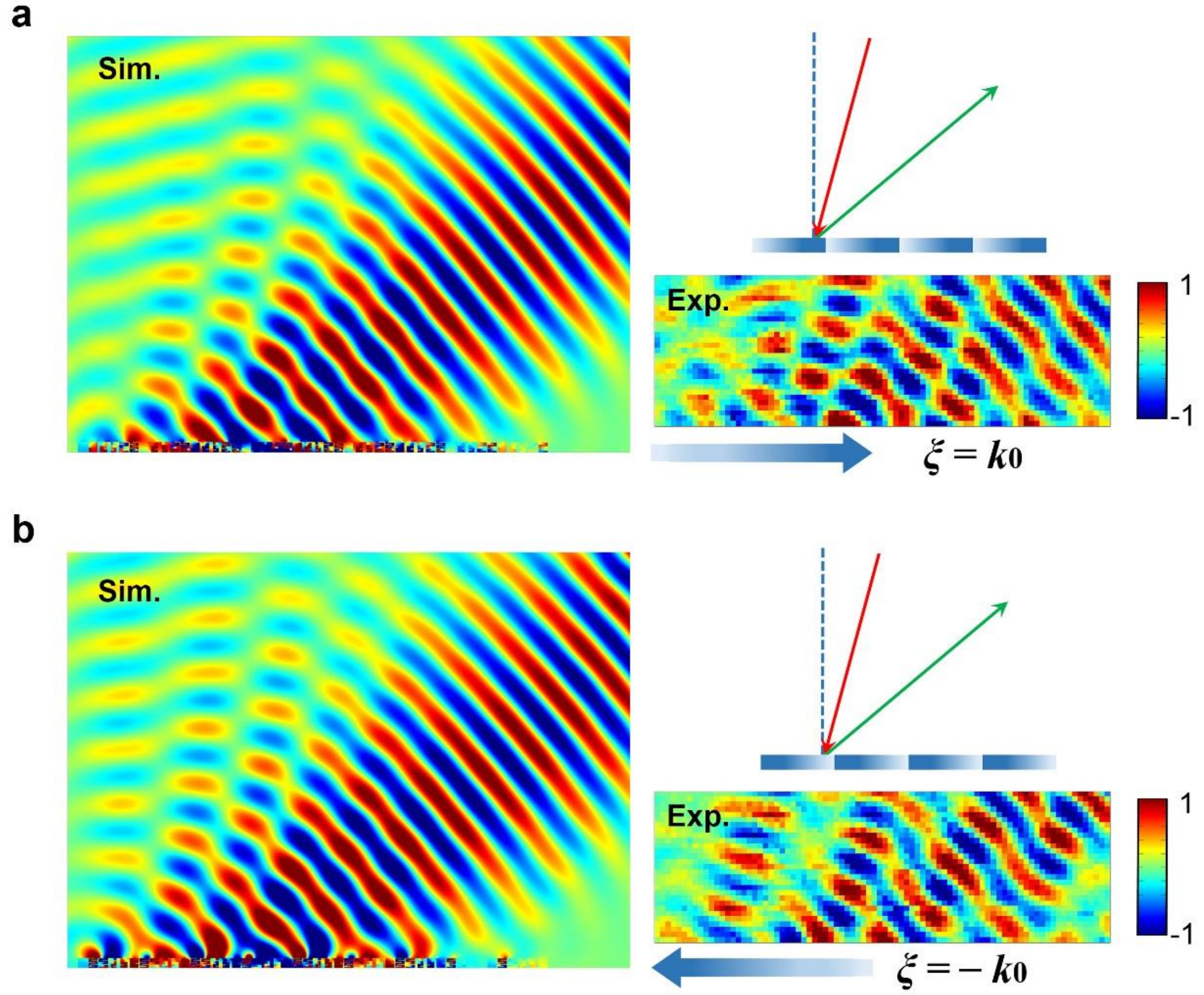

**Figure 5 Experiment verification of the all-angle negative reflection.** (a)Simulation and experiment measurement of the scattered field when the metasurface having $k_0$ or (b) $-k_0$ surface phase gradient is illuminated at $-15^\circ$ incidence.

Figure 5(a) illustrates the calculated and measured scattered acoustic field when the metasurface with $k_0$ surface phase gradient is illuminated under $-15^\circ$ incidence, the measured reflected angle is $49.9^\circ$, which is close to the theoretical value $49.2^\circ$. Then we inverse the direction of surface phase gradient without varying the incident angle. According to the generalized law of reflection, there exists no free space scattered fields, however, both theoretical calculation and experimental measurements prove the existence of the negative reflection beyond the critical angle incidence, the measured reflected angle is $48.8^\circ$, see Figure 5(b). The blue hollow triangles depicted in Figure 4(e) are all three set of measured data following the same measuring procedure, which agree well with theoretical and simulation results. Therefore, the all-angle negative reflection can be realized with acoustic gradient metasurface.

Figure 6 demonstrates the scattered properties of the acoustic gradient metasurface while the beam incidents at $-10^{\circ}$ and $-20^{\circ}$, respectively. It can be seen from Figure 6(c) and (e) that when the beam is incident at small angle, the 0-th diffraction order which corresponds to the specular reflection would appear. It should be noted that the scattered field depicted in Figure 6(f) exhibiting partial interference pattern is caused by the reflection of the speaker array, and that is why we choose relative small incident angle (no greater than $20^{\circ}$ in this work) to conduct the experimental measurement.

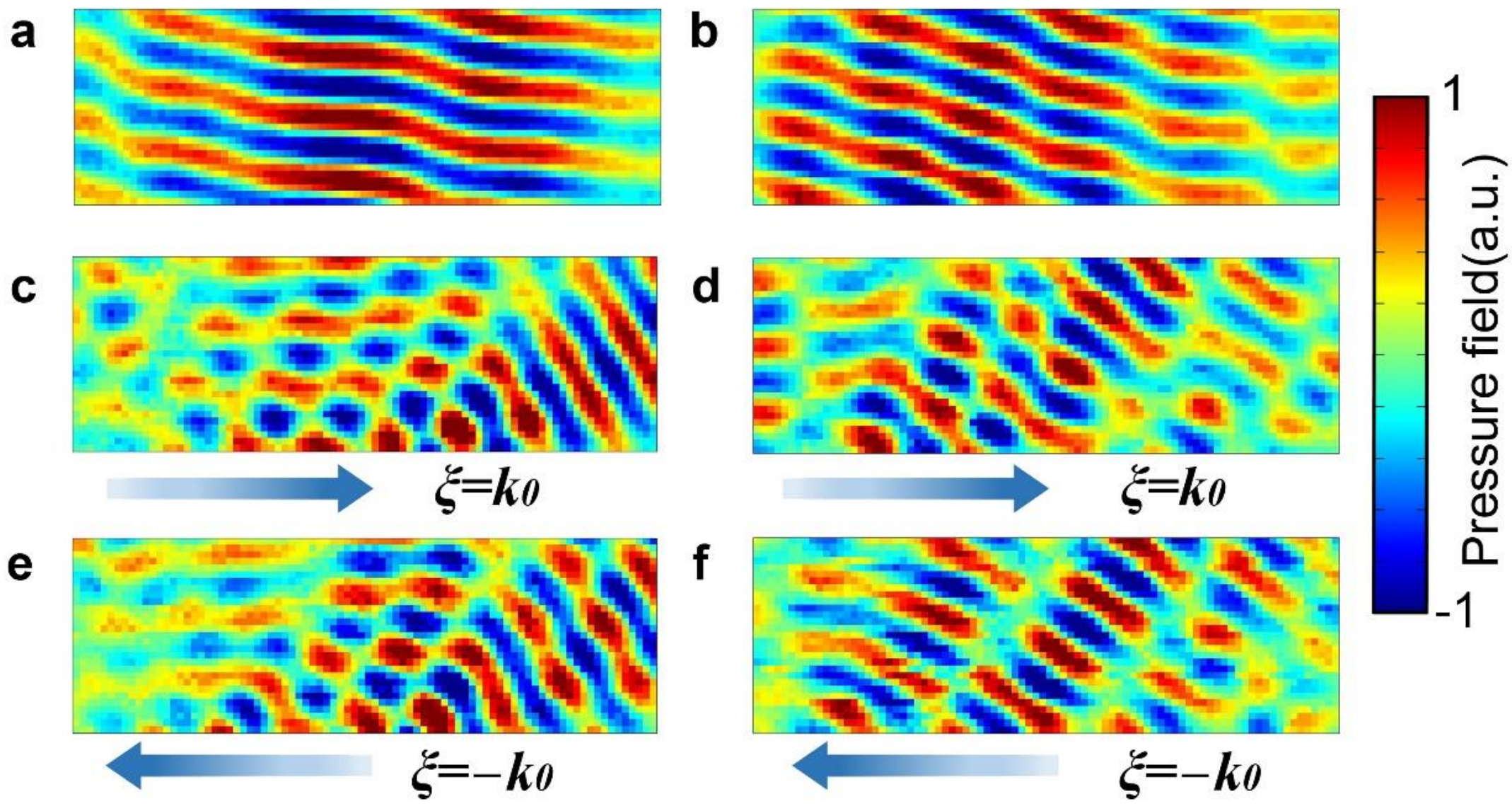

**Figure 6 Measured background field and scattered field when AGM being illuminated at other incident angles.** (a) The measured background field(without the metasurface) incident at $-10^{\circ}$ and (b) $-20^{\circ}$. (c) The measured scattered field of the metasurface with $k_0$ surface phase gradient being illuminated at $-10^{\circ}$ and $-20^{\circ}$ reflects at $56.3^{\circ}$ and (d) $40^{\circ}$ respectively. (e) The measured scattered field of the metasurface with $-k_0$ surface phase gradient being illuminated at $-10^{\circ}$ and $-20^{\circ}$ reflects at $58.4^{\circ}$ and (f) $41.6^{\circ}$ correspondingly.

## Discussion

The all-angle negative reflection characteristic possessed by the acoustic gradient metasurface has the potential application in composing ultrathin compact acoustic insulator operating in reflection mode[34], for the incident acoustic wave would be reflected back into the incident half space and thereby avoid the cross talk between the acoustic waves coming from opposite directions. Moreover, the reflected acoustic gradient metasurface studied in our work also shows good beam deflection functionality over 700 Hz bandwidth, which shows good promise for constructing broadband acoustic back reflector, see Figure 7. Different from the negative reflection phenomena which has been theoretically predicted to exist in strong chiral materials[43], the all-angle negative reflection obtained in acoustic gradient metasurface involves the selective excitation of the waveguide mode in the form of Floquet-Bloch waves and the Bragg scattering contributed by the supercell lattice. It should be noted that when the amplitude of surface phase gradient is greater or less than $k_0$, more complicated wavefront manipulation functionality can be obtained with the acoustic gradient metasurface[35]. Beyond the critical angle illumination, higher order negative Floquet-Bloch modes could be excited and couple the energy into other diffraction orders. Furthermore, the critical angle indeed corresponds to the Bragg scattering point of the gradient metasurface. Therefore, several diffraction orders, such as specular reflection and negative reflection, would appear simultaneously around the critical angle.

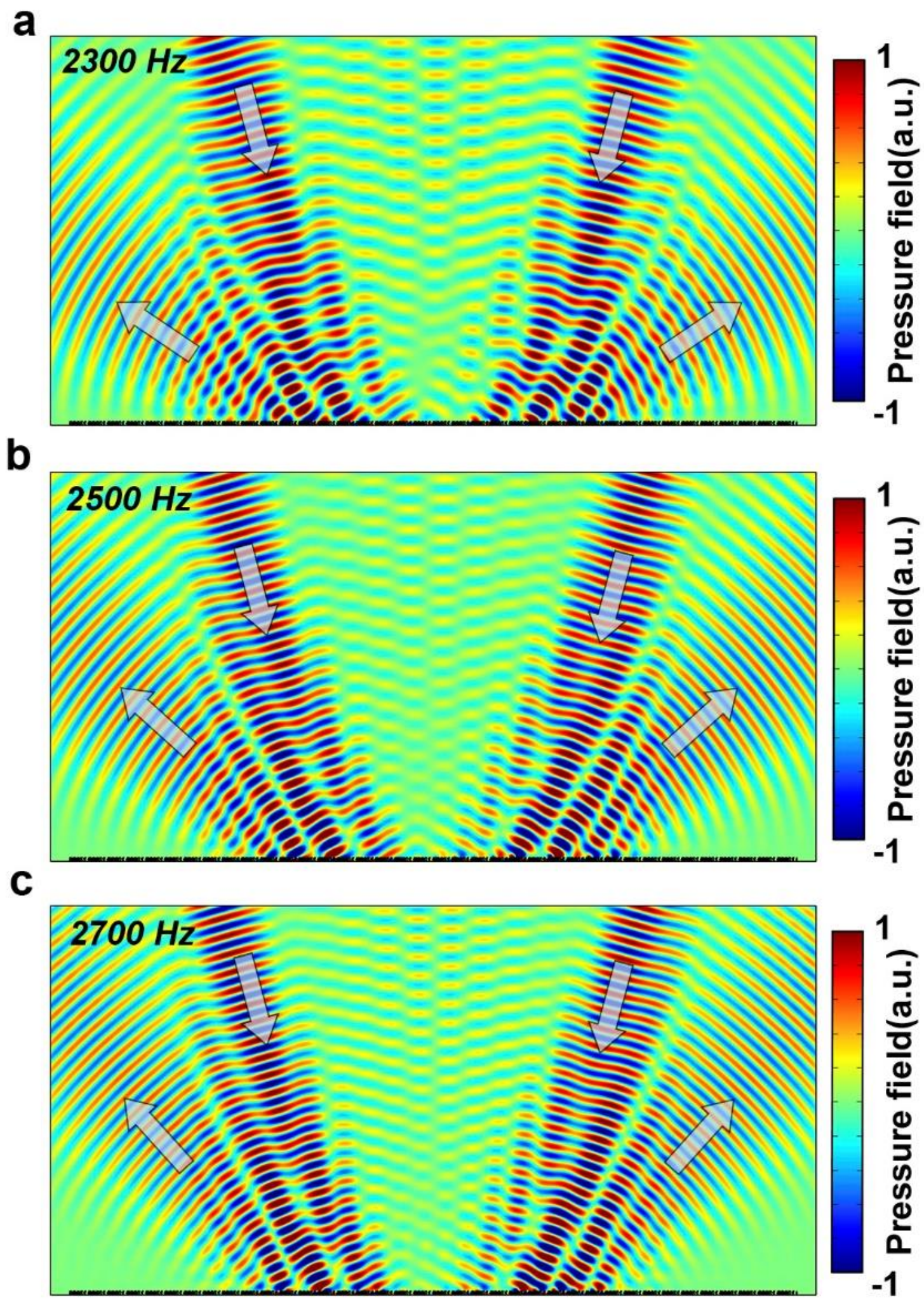

**Figure 7 Broadband frequency response of the acoustic gradient metasurface.** Acoustic back reflector operates at (a) 2300 Hz, (b) 2500 Hz and (c) 2700 Hz. Two beams are incident onto the acoustic gradient metasurface at $\pm 15^{\circ}$ simultaneously.

In summary, we theoretically and experimentally study the all-angle negative reflection phenomenon realized by the acoustic gradient metasurface. The all-angle negative reflection has been perfectly verified by experimental measurements which agree well with theoretical calculations. The Floquet-Bloch mode interpretation based on the mode expansion theory has been utilized to elucidate the underlying mechanism of this nontrivial phenomenon, which indicates that the all-angle negative reflection can be realized only if the dominated first negative Floquet-Bloch mode is efficiently excited. Moreover, we can predict more complicated beam deflection phenomena while the amplitude of surface phase gradient is different from $k_0$, which are attributed to the higher order negative Floquet-Bloch modes. Generally speaking, our work can also be extended to the transmission case and offers a new solution for wavefront manipulation.

## Methods

**Simulations.** In this paper, all full-wave simulations are performed with commercial finite-element solver COMSOL Multiphysics. The sound speed is chosen as 340 m/s, and the metasurface structure is assumed to be sound hard by applying the interior sound hard boundary condition at the air-structure interface. The acoustic field are all normalized by the background field of which the amplitude is set to be 1 Pa for the simulation. The reflected angle can be obtained from the far field distribution of the scattered acoustic field.

**Data availability.** The data that support the conclusions of this work are available from the corresponding author on request.

## Acknowledgements

Thanks for the kind help of Ms. Yuhong Na in preparing the high quality figures. This work was supported by National Natural Science Foundation of China (NSFC) (50836002, 51176041).

## Author contributions

B. Liu conceived the original idea and designed the experiments; B. Liu and J. Zhao performed the theoretical and numerical calculations; B. Liu, X. Xu and W. Zhao conducted the experiments; B. Liu, J. Zhao and W. Zhao prepared the manuscript; Y. Jiang supervised the project.

## Additional information

**Supplementary information** accompanies this paper.

**Competing financial interests:** The authors declare no competing financial interests.

## References

1. Yu, N. *et al.* Light propagation with phase discontinuities: Generalized laws of reflection and refraction, *Science* **334**, 333 (2011).
2. Zhang, L., Mei, S., Huang, K. & Qiu, C. Advances in full control of electromagnetic waves with metasurfaces. *Adv. Opt. Mater*. **4**, 818 (2016).
3. Lin, D., Fan, P., Hasman, E. & Brongersma, M. L. Dielectric gradient metasurface optical elements. *Science* **345**, 298 (2014).
4. Khorasaninejad, M. *et al*. Metalenses at visible wavelengths: Diffraction-limited focusing and subwavelength resolution imaging. *Science* **352**, 1190 (2016).
5. Huang, L. *et al*. Three dimensional optical holography using a plasmonic metasurface. *Nat. Commun*. **4**, 2808 (2013).
6. Zheng, G. *et al*. Metasurface holograms reaching 80% efficiency. *Nat. Nanotechnol*. **10**, 308 (2015).
7. Ni, X., Wong, Z., Mrejen, M., Wang, Y. & Zhang, X. An ultrathin invibility skin cloak for visible light. *Science* **349**, 1310 (2015).
8. Kamali, S. M., Arbabi, A., Arbabi, E., Horie, Y. & Faraon, A. Decoupling optical function and geometrical form using conformal flexible dielectric metasurfaces. *Nat. Commun*. **7**, 11618 (2016).
9. Cheng, J., Jafar-Zanjani, S. & Mosallaei, H. All dielectric ultrathin conformal metasurfaces: lensing and cloaking applications at 532 nm wavelength. *Sci. Rep*. **6**, 38440 (2016).
10. Li, Y., Liang, B., Gu, Z., Zou, X. & Cheng, J. Reflected wavefront manipulation based on ultrathin planar acoustic metasurfaces. *Sci. Rep*. **3**, 2546 (2013).
11. Tang, K. *et al*. Anomalous refraction of airborne sound through ultrathin metasurfaces. *Sci. Rep*. **4**, 6517 (2014).
12. Xie, Y. *et al*. Wavefront modulation and subwavelength diffractive acoustics with an acoustic metasurface. *Nat. Commun*. **5**, 5553 (2014).
13. Liang, Z. & Li, J. Extreme acoustic metamaterial by coiling up space. *Phys. Rev. Lett*. **108**, 114301 (2012).
14. Xie, Y., Popa, B.-I., Zigoneanu, L. & Cummer, S. A. Measurement of a broadband

negative index with space-coiling acoustic metamaterials. *Phys. Rev. Lett*. **110**, 175501 (2013).

15. Xie, Y., Konneker, A., Popa, B.-I. & Cummer, S. A. Tapered labyrinthine acoustic metamaterials for broadband impedance matching. *Appl. Phys. Lett*. **103**, 201906 (2013).
16. Li, Y. *et al*. Acoustic focusing by coiling up space. *Appl. Phys. Lett*. **101**, 233508 (2012).
17. Zhu, X. *et al*. Implementation of dispersion-free slow acoustic wave propagation and phase engineering with helical-structured metamaterials. *Nat. Commun*. **7**, 11731 (2016).
18. Zhao, J., Li, B., Chen, Z. & Qiu, C. Redirection of sound waves using acoustic metasurface. *Appl. Phys. Lett*. **103**, 151604 (2013).
19. Zhao, J., Li, B., Chen, Z. & Qiu, C. Manipulating acoustic wavefront by inhomogeneous impedance and steerable extraordinary reflection. *Sci. Rep*. **3**, 2537 (2013).
20. Zhai, S. *et al*. Manipulation of transmitted wavefront using planar acoustic metasurfaces. *Appl. Phys. A* **120**, 1283 (2015).
21. Ding, C., Zhao, X., Chen, H., Zhai, S. & Shen, F. Reflected wavefronts modulation with acoustic metasurface based on double-slit hollow sphere. *Appl. Phys. A* 120, 487 (2015).
22. Li, Y., Jiang, X., Liang, B., Cheng, J. & Zhang, L. Metascreen-based acoustic passive phased array. *Phys. Rev. Appl*. **4**, 024003 (2015).
23. Wang, W., Xie, Y., Konneker, A., Popa, B.-I. & Cummer, S. A. Design and demonstration of broadband thin planar diffractive acoustic lenses. *Appl. Phys. Lett*. **105**, 101904 (2014).
24. Li, Y. *et al*. Experimental realization of full control of reflected waves with subwavelength acoustic metasurfaces. *Phys. Rev. Appl*. **2**, 064002 (2014).
25. Jiang, X., Li, Y., Liang, B., Cheng, J. & Zhang, L. Convert acoustic resonances to orbital angular momentum. *Phys. Rev. Lett*. **117**, 034301 (2016).
26. Ye, L. *et al*. Making sound vortices by metasurfaces. *AIP Adv*. **6**, 085007 (2016).

27. Ma, G., Yang, M., Xiao, S., Yang, Z. & Sheng, P. Acoustic metasurface with hybrid resonances. *Nat. Mater*. **13**, 873 (2014).
28. Li, Y. & Assouar, M B. Acoustic metasurface-based perfect absorber with deep subwavelength thickness. *Appl. Phys. Lett*. **108**, 063502 (2016).
29. Faure, C., Richoux, O., Felix, S. & Pagneux, V. Experiments on metasurface carpet cloaking for audible acoustics. *Appl. Phys. Lett*. **108**, 064103 (2016)
30. Esfahlani, H., Karkar, S. & Lissek, H. Acoustic carpet cloaking based on an ultrathin metasurface. *Phys. Rev. B* **94**, 014302 (2016).
31. Sun, S. *et al*. Gradient-index meta-surfaces as a bridge linking the propagating waves and surface waves. *Nat. Mater*. **11**, 426 (2012).
32. Sun, S. *et al*. High-efficiency broadband anomalous reflection by gradient meta-surfaces. *Nano Lett*. **12**, 6223 (2012).
33. Aieta, F. *et al*. Out-of-plane reflection and refraction of light by anisotropic optical antenna metasurfaces with phase discontinuities. *Nano Lett*. **12**, 1702 (2012).
34. Liu, B., Zhao, W. & Jiang, Y. Full-angle negative reflection realized by a gradient acoustic metasurface. *AIP Adv*. **6**, 115110 (2016).
35. Liu, B., Zhao, W. & Jiang, Y. Apparent negative reflection with the gradient acoustic metasurface by integrating supercell periodicity into the generalized law of reflection. *Sci. Rep*. **6**, 38314 (2016).
36. Li, Y., Qi, S. & Assouar, M B. Theory of metascreen-based acoustic passive phased array. *New J. Phys*. **18**, 043024 (2016).
37. Xu, Y., Fu, Y. & Chen, H. Steering light by a sub-wavelength metallic grating from transformation optics. *Sci. Rep*. **5**, 12219 (2015).
38. Xiao, S. *et al*. Mode-expansion theory for inhomogeneous meta-surfaces. *Opt. Express* **21**, 27219 (2013).
39. Xie, Y., Zakharian, A. R., Moloney, J. V. & Mansuripur, M. Transmission of light through periodic arrays of sub-wavelength slits in metallic hosts. *Opt. Express* **14**, 6400 (2006).
40. Xie, Y., Zakharian, A. R., Moloney, J. V. & Mansuripur, M. Optical transmission at oblique incidence through a periodic array of sub-wavelength slits in a metallic host.

*Opt. Express* **14**, 10220 (2006).

41. Lalanne, P., Hugonin, J. P. & Chavel, P. Optical properties of deep Lamellar gratings: A coupled Bloch-mode perspective. *J. Lightwave Technol*. 24, 2442 (2006).
42. Zhao, D., Ye, Y., Xu, S., Zhu, X. & Yi, L. Broadband and wide-angle negative reflection at a photonic crystal boundary. *Appl. Phys. Lett*. **104**, 043503 (2014).
43. Zhang, C. & Cui, T. Negative reflections of electromagnetic waves in a strong chiral medium. *Appl. Phys. Lett*. **91**, 194101 (2007).